\documentclass[twocolumn,english,aps,pra,superscriptaddress,twocolumn,address,showpacs,showacknowledgments]{revtex4}
\pdfoutput=1
\usepackage[T1]{fontenc}
\usepackage[latin9]{inputenc}
\usepackage{color}
\usepackage{babel}
\usepackage{amsmath}
\usepackage{amssymb}
\usepackage{graphicx}
\usepackage{esint}
\usepackage[unicode=true,
 bookmarks=true,bookmarksnumbered=false,bookmarksopen=false,
 breaklinks=false,pdfborder={0 0 1},backref=false,colorlinks=false]
 {hyperref}
\hypersetup{pdftitle={Full photon statistics of a light beam transmitted through an optomechanical system},
 pdfauthor={Andreas Kronwald, Max Ludwig and Florian Marquardt},
 pdfkeywords={Optomechanics,Ultrastrong Coupling, Photon Statistics, Quantum Jump Trajectories,Bunching,Antibunching}}
\makeatletter
\@ifundefined{textcolor}{}
{%
 \definecolor{BLACK}{gray}{0}
 \definecolor{WHITE}{gray}{1}
 \definecolor{RED}{rgb}{1,0,0}
 \definecolor{GREEN}{rgb}{0,1,0}
 \definecolor{BLUE}{rgb}{0,0,1}
 \definecolor{CYAN}{cmyk}{1,0,0,0}
 \definecolor{MAGENTA}{cmyk}{0,1,0,0}
 \definecolor{YELLOW}{cmyk}{0,0,1,0}
 }

\usepackage{babel}\usepackage{amstext}
\usepackage{xcolor}

\@ifundefined{textcolor}{}{%
 \definecolor{BLACK}{gray}{0}
 \definecolor{WHITE}{gray}{1}
 \definecolor{RED}{rgb}{1,0,0}
 \definecolor{GREEN}{rgb}{0,1,0}
 \definecolor{BLUE}{rgb}{0,0,1}
 \definecolor{CYAN}{cmyk}{1,0,0,0}
 \definecolor{MAGENTA}{cmyk}{0,1,0,0}
 \definecolor{YELLOW}{cmyk}{0,0,1,0}
 }

\makeatother

\begin{document}

\title{Full photon statistics of a light beam transmitted through an optomechanical
system}

\author{Andreas Kronwald}

\email{andreas.kronwald@physik.uni-erlangen.de}

\selectlanguage{english}%

\affiliation{Friedrich-Alexander-Universität Erlangen-Nürnberg, Staudtstr. 7,
D-91058 Erlangen, Germany}

\author{Max Ludwig}

\affiliation{Friedrich-Alexander-Universität Erlangen-Nürnberg, Staudtstr. 7,
D-91058 Erlangen, Germany}

\author{Florian Marquardt}

\affiliation{Friedrich-Alexander-Universität Erlangen-Nürnberg, Staudtstr. 7,
D-91058 Erlangen, Germany}

\affiliation{Max Planck Institute for the Science of Light, Günther-Scharowsky-Straße
1/Bau 24, D-91058 Erlangen, Germany}

\begin{abstract}
In this paper, we study the full statistics of photons transmitted
through an optical cavity coupled to nanomechanical motion. We analyze
the entire temporal evolution of the photon correlations, the Fano
factor, and the effects of strong laser driving, all of which show
pronounced features connected to the mechanical backaction. In the
regime of single-photon strong coupling, this allows us to predict
a transition from sub-Poissonian to super-Poissonian statistics for
larger observation time intervals. Furthermore, we predict cascades
of transmitted photons triggered by multi-photon transitions. In this
regime, we observe Fano factors that are drastically enhanced due
to the mechanical motion.
\end{abstract}
\maketitle

\section{Introduction}
\begin{figure}[t]
\includegraphics{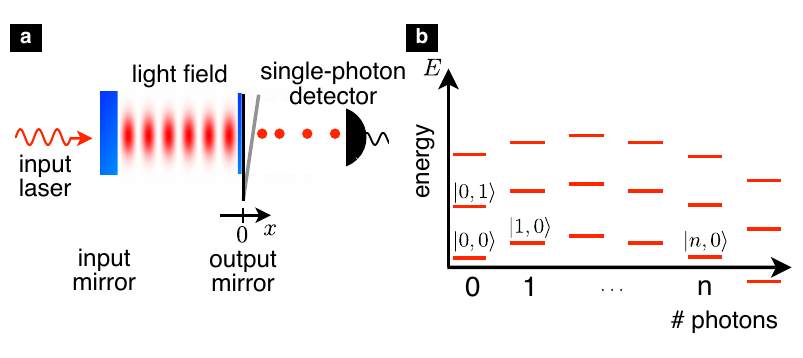}\caption{(a) The standard optomechanical setup, where transmitted photons are
detected by a single-photon detector. As we will show, the strong
backaction of the mechanical motion onto the light field dynamics
leads to new features in the photon correlations. (b) Nonlinear level
scheme induced by the light-mechanics coupling. The eigenstates read
$|n_{a},n_{b}\rangle$, where $n_{a}$ $(n_{b})$ denotes the photon
(phonon) number.}
\label{fig:Standard_Setup_and_Level_Scheme} 
\end{figure}

Photon statistics serves as a powerful tool in quantum optical experiments.
For example, it has been used to identify the strong coupling between
radiation and an atom by observing non-classical photon-antibunching
\cite{BirnbaumPhotonBlockade2005} and multi-photon transitions \cite{KubanekTwo_Photon_Gateway2008,MajumdarProbingDressedStates2012,ReinhardStronglyCorrelatedPhotons2012}.
While all of these examples show how photon statistics can be influenced
by coupling to internal degrees of freedom, only a few experiments
have demonstrated the effects of any kind of mechanical motion. These
mostly involve the classical motion of atoms flying through cavities
\cite{MuenstermannSingleAtomDynamics1999,TerracianoPhotonBurst2009,HoodCavityMicroscope2000,DohertySingleAtomTrapping2000}.
A first step into the quantum domain has been reported in \cite{RotterMonitoringSingleIonsMotion2008}
by showing the impact of quantized ionic motion onto the photon statistics.
In all of these examples, however, the statistics is heavily influenced
by both internal and external degrees of freedom. We will analyze
a setup where, in contrast to those systems, the photon correlations
(both on short and long time scales) are entirely due to the coupling
of photons to a \textit{single} degree of freedom, namely a quantum
mechanical vibrating resonator. 

\textcolor{black}{These cavity optomechanical systems \cite{KippenbergCavity2008}
offer a new domain to study the effects of quantized mechanical motion
on the statistics of photons, without the need for atom trapping.
The field has been developing rapidly during the past few years. It
is driven by the goals of probing the quantum motion of nanomechanical
devices, implementing ultrasensitive measurements, and exploiting
the light-mechanics coupling for quantum information processing, as
well as answering fundamental questions about quantum mechanics. Recently,
the mechanical degree of freedom was cooled down close to its quantum
ground state using optomechanical sideband cooling \cite{TeufelSideband2011,ChanLaser2011}.
Thus, it now becomes important to analyze probes for the quantum dynamics
of these systems. Experimentally, the most straightforward access
is provided by the light field. }

\textcolor{black}{Earlier works \cite{FabreNoiseReduction1994,ManciniNoiseReduction1994}
studied the linearized dynamics of small fluctuations in the context
of homodyne detection, where the light field was described by continuous
variables and squeezing effects were predicted. A particularly powerful
tool is given by the statistics of single photons, as those are very
sensitive to interactions and reveal the temporal correlations produced
by the dynamics. A first study in this direction recently led to the
prediction of photon blockade \cite{RablPhoton2011} in optomechanical
systems. Here, we will take a significant step further by analyzing
the full statistics of the stream of transmitted photons. In contrast
to \cite{RablPhoton2011}, our numerical approach allows us to study
many new aspects, including the regime of strong laser driving, the
full temporal structure of photon correlations, the Fano factor, and
higher moments of the distribution of transmitted photons. }

\textcolor{black}{All of these properties become particularly interesting
in the regime of single-photon strong light-mechanics coupling, which
(although challenging) is now being approached experimentally \cite{MurchObservation2008,BrenneckeCavity2008,TeufelSideband2011,ChanLaser2011,VerhagenQuantum-coherent2012,ChanOptimizedOptomechanicalCrystal2012}.
In this regime, the single photon coupling rate becomes comparable
to the photon decay rate, and a few remarkable phenomena have already
been predicted. A {}``classical to quantum crossover'' may be observed
in nonlinear optomechanical dynamics \cite{Ludwigoptomechanical2008},
and photon blockade effects \cite{RablPhoton2011}, non-Gaussian \cite{NunnenkampSingle-Photon2011}}\textbf{\textcolor{black}{{}
}}\textcolor{black}{and nonclassical \cite{QianQuantum2011} mechanical
states may be produced. Additionally, diverse Schrödinger-cat states
may be generated \cite{ManciniPonderomotive1997,BosePreparation1997},
multiple cooling resonances may be observed \cite{NunnenkampMultipleCoolingResonance2012},
and certain dark states may exist \cite{XuDarkStates2012}. Furthermore,
QND photon and phonon readout and nonlinearities could be enhanced
in appropriate two-mode setups \cite{LudwigOptomechanical2012,StannigelOptomechanical2012}.}

\textcolor{black}{This paper is organized as follows: First, we introduce
the optomechanical model in Sec.~\ref{sec:The-Model}, and show how
photons and phonons can be decoupled formally. In Sec.~\ref{sec:Methods},
we define the Lindblad master equation for the optomechanical system
to describe its dissipative dynamics. Then we discuss the quantum
jump trajectory technique giving access to the full counting statistics,
which is described thereafter. Furthermore, we introduce the Fano
factor as a measure for the photon statistics and connect it to the
two-photon correlation function $g^{(2)}(\tau)$. At the end of this
section, we show how the Fano factor is influenced by finite detector
efficiencies. In Sec. \ref{sub:Weak-Laser-Drive} we discuss a weakly
driven, strongly coupled optomechanical system, with emphasis on the
full time-dependence of the photon correlator. In Sec. \ref{sub:Cascade-of-Photon},
we present our main results in the regime of strong driving, where
we introduce and discuss the regime of photon cascades.}

\section{The Model\label{sec:The-Model}}
We consider a generic optomechanical setup consisting of a laser-driven
optical cavity, where one of the end-mirrors is attached to a vibrating
resonator, see Fig.~\ref{fig:Standard_Setup_and_Level_Scheme}a.
The coherent dynamics is described by the Hamiltonian 
\begin{equation}
\hat{H}=-\hbar\Delta\hat{a}^{\dagger}\hat{a}-\hbar g_{0}\left(\hat{b}^{\dagger}+\hat{b}\right)\hat{a}^{\dagger}\hat{a}+\hbar\Omega\hat{b}^{\dagger}\hat{b}+\hbar\alpha_{L}\left(\hat{a}^{\dagger}+\hat{a}\right)\,,\label{eq:standard_optomechanical_hamiltonian}
\end{equation}
written in a frame rotating at the laser frequency $\omega_{L}$ \cite{LawHamiltonian1997,KippenbergCavity2008}.
Here, $\hat{a}$ ($\hat{b}$) is the photon (phonon) annihilation
operator, $\Delta=\omega_{L}-\omega_{\text{cav}}$ is the laser detuning
from resonance, and $g_{0}$ is the single-photon optomechanical coupling
constant. $\Omega$ denotes the frequency of the mechanical resonator,
and $\alpha_{L}$ is the laser driving amplitude. 

Photons and phonons can be decoupled by applying the so-called polaron
transformation \cite{ManciniPonderomotive1997,BosePreparation1997}
$\hat{\tilde{H}}=\hat{U}^{\dagger}\hat{H}\hat{U}$ to Eq.~(\ref{eq:standard_optomechanical_hamiltonian}),
where $\hat{U}=\exp\left[g_{0}\hat{a}^{\dagger}\hat{a}(\hat{b}^{\dagger}-\hat{b})/\Omega\right]$.
This yields
\begin{equation}
\hat{\tilde{H}}=-\hbar\Delta\hat{a}^{\dagger}\hat{a}+\hbar\Omega\hat{b}^{\dagger}\hat{b}-\hbar\frac{g_{0}^{2}}{\Omega}\left(\hat{a}^{\dagger}\hat{a}\right)^{2}+\hbar\alpha_{L}\left(\mbox{\ensuremath{\hat{a}^{\dagger}\hat{D}^{\dagger}}+h.c.}\right)\,,\label{eq:polaron_transformed_hamiltonian}
\end{equation}
where the mechanical displacement operator $\hat{D}=\exp\left[g_{0}(\hat{b}^{\dagger}-\hat{b})/\Omega\right]$
now enters the transformed driving term. The elimination of the photon-phonon
coupling gives rise to a photon-photon interaction term. This leads
to an anharmonic photon energy level scheme, cf. Fig.~\ref{fig:Standard_Setup_and_Level_Scheme}b,
which drastically influences the intracavity photon statistics if
$g_{0}^{2}\gtrsim\kappa\Omega$ \cite{RablPhoton2011}.

\section{Methods\label{sec:Methods}}
\subsection{Dissipative dynamics}
The dissipative dynamics of the open optomechanical system is described
by a Lindblad master equation
\begin{align}
\dot{\hat{\rho}}= & \frac{\mathrm{i}}{\hbar}\left[\hat{\rho},\hat{H}\right]+\kappa_{I}\mathcal{D}[\hat{a}]\hat{\rho}+\Gamma_{M}\left(n_{\text{th}}+1\right)\mathcal{D}[\hat{b}]\hat{\rho}\nonumber \\
 & +\Gamma_{M}n_{\text{th}}\mathcal{D}\left[\hat{b}^{\dagger}\right]\hat{\rho}-\frac{\kappa_{O}}{2}\left\{ \hat{a}^{\dagger}\hat{a},\hat{\rho}\right\} \nonumber \\
 & +\kappa_{O}\,\hat{a}\hat{\rho}\hat{a}^{\dagger}\,,\label{eq:LME}
\end{align}
where $\hat{\rho}$ is the density matrix for the photons and phonons.
$\kappa_{I}$ ($\kappa_{O}$) denotes the photon loss rate through
the input (output) mirror, where $\kappa=\kappa_{I}+\kappa_{O}$ is
the total cavity decay rate and $\Gamma_{M}$ is the phonon decay
rate. Furthermore, $n_{\text{th}}^{-1}=\exp\left(\hbar\Omega/k_{B}T\right)-1$,
where $T$ is the mechanical bath temperature. The Lindblad superoperator
reads $\mathcal{D}[\hat{a}]\hat{\rho}=\hat{a}\hat{\rho}\hat{a}^{\dagger}-\left\{ \hat{a}^{\dagger}\hat{a},\hat{\rho}/2\right\} $,
where $\left\{ \hat{a},\hat{b}\right\} =\hat{a}\hat{b}+\hat{b}\hat{a}$.
\subsection{Quantum jump trajectories\label{sub:Quantum-Jump-Trajectories}}
In experiments, the photon statistics can be analyzed by measuring
the times when a photon leaves the cavity through the output mirror,
cf. Fig. \ref{fig:Standard_Setup_and_Level_Scheme}a. To simulate
this detection process, we use quantum jump trajectories \cite{DalibardWave-function1992,GardinerWave-function1992,CarmichaelOpen1993,Plenioquantum-jump1998,WisemanInterpretation1993},
which can be understood as an \textquotedblleft{}unraveling\textquotedblright{}
of the master equation (\ref{eq:LME}). It has recently been employed
to discuss single-phonon detection in optomechanical systems \cite{GangatPhonon2011}.

If a photon is detected ({}``photon jump'') during a time interval
$[t,t+\delta t]$ with probability $p_{j}=\eta\kappa_{O}\langle\hat{a}^{\dagger}\hat{a}\rangle(t)\delta t$,
the information about the system is updated according to 
\begin{equation}
\hat{\rho}(t+\delta t)=\frac{\hat{a}\hat{\rho}(t)\hat{a}^{\dagger}}{\mathrm{Tr}(\hat{a}^{\dagger}\hat{a}\hat{\rho}(t))}\,.\label{eq:photon_jump_evolution}
\end{equation}
Here, $\eta$ is the photon detector efficiency, i.e. the ratio of
the number of detected photons to the total number of transmitted
photons.

If no photon is detected, $\hat{\rho}$ evolves according to Eq.~(\ref{eq:LME})
(where the original Hamiltonian (\ref{eq:standard_optomechanical_hamiltonian}),
without polaron transformation, is used in our numerics), with the
last term of Eq.~(\ref{eq:LME}) replaced by $(1-\eta)\kappa_{O}\hat{a}\hat{\rho}\hat{a}^{\dagger}$.
After each such time step, the state has to be normalized again. This
evolution describes the increase of our knowledge due to the absence
of a detection event \cite{UedaQuantum1990,CarmichaelOpen1993}. 

Overall, we obtain stochastic traces of photon detection events which
directly correspond to what would be observed in an experiment that
registers single photons. This is one of the principal advantages
of employing the quantum jump trajectory method. From these trajectories,
we gain access to the full statistics, including arbitrary moments
and the complete time-dependence of photon correlations.
\subsection{Full counting statistics\label{sub:Full-Counting-Statistics}}
The full counting statistics $p(N,T_{S})$ \cite{MandelOptical1995}
is the probability of measuring $N$ transmitted photons in a time
interval $T_{S}$, for a constant detection-rate $\dot{\bar{N}}=\eta\kappa_{O}\,\bar{n}$.
Here, $\bar{n}$ is the steady state photon number of Eq. (\ref{eq:LME}).
The fluctuations of the detected photon number can be characterized
via the Fano factor 
\begin{equation}
\mathcal{F}_{\text{c}}(T_{S})=\left(\left\langle N^{2}\right\rangle -\left\langle N\right\rangle ^{2}\right)/\left\langle N\right\rangle \,,
\end{equation}
where $\left\langle N^{m}\right\rangle =\sum_{N=0}^{\infty}N^{m}\, p(N,T_{S})$.
If $\mathcal{F}_{\text{c}}(T_{S})<1$ there is sub-Poissonian statistics
(on that observation time-scale), whereas for $\mathcal{F}_{\text{c}}(T_{S})>1$
the photons obey super-Poissonian statistics. 

The long-time limit of the Fano factor yields the zero-frequency shot
noise power, $S_{II}[\omega=0]=\dot{\bar{N}}\mathcal{F}_{{\rm c}}(\infty)$$ $.
Thus, $\mathcal{F}_{c}(\infty)$ can be interpreted as the effective
number of detected photons during a single transmission process (i.e.
during a single {}``bunch'' of photons).
\subsection{Photon correlations $g^{(2)}(\tau)$ and connection to the Fano factor\label{sub:PConnection_of_g2_and_Fano_factor}}
\textcolor{black}{Another measure of photon correlations is given
by the two-photon correlation function
\begin{equation}
g^{(2)}(\tau)=\frac{\left\langle \hat{a}^{\dagger}(0)\hat{a}^{\dagger}(\tau)\hat{a}(\tau)\hat{a}(0)\right\rangle }{\left\langle \hat{a}^{\dagger}(0)\hat{a}(0)\right\rangle ^{2}}\,.\label{eq:def_g2_tau}
\end{equation}
Here we have assumed that a steady state exists, and all expectation
values in this definition are taken with respect to the steady state
density matrix of the full Lindblad master equation (\ref{eq:LME}). }

\textcolor{black}{The two-photon correlation function $g^{(2)}(\tau)$
can be interpreted as a measure for detecting a photon at time $\tau$
conditioned on the detection of a photon at time $\tau=0$. Using
$g^{(2)}(\tau)$ we can distinguish three regimes \cite{WallsMilburn1994}: }
\begin{enumerate}
\item \textcolor{black}{If $g^{(2)}(\tau)=1$ for all delay times $\tau$,
the photon detection events are statistically independent and obey
Poissonian statistics. }
\item \textcolor{black}{Photon bunching is defined via $g^{(2)}(0)>1$.
In this case, there is a tendency that photons arrive in groups.}
\item \textcolor{black}{The photons are anti-bunched when $g^{(2)}(0)<1$,
which is a pure quantum effect. In contrast to photon bunching, the
transmitted photons tend to avoid each other.}
\end{enumerate}
The Fano factor $\mathcal{F}_{\text{c}}$ defined in Sec. \ref{sub:Full-Counting-Statistics}
is connected to this two-photon correlation function $g^{(2)}(\tau)$
via \cite{MandelOptical1995} 
\begin{equation}
\mathcal{F}_{\text{c}}(T_{S})=1+\dot{\bar{N}}\int_{-T_{S}}^{T_{S}}\mathrm{d}\tau\,\left(g^{(2)}(\left|\tau\right|)-1\right)\left(1-|\tau|/T_{S}\right)\,.\label{eq:fano_factor_small_sampling_times}
\end{equation}
\textcolor{black}{Thus, for small sampling times $T_{S}$, we can
distinguish photon anti-bunching and bunching using $\mathcal{F}_{c}(T_{S})$:
If the photons are anti-bunched, $g^{(2)}(0)<1$ and, hence, the slope
of the Fano factor $\mathcal{F}_{c}(T_{S})$ is negative for small
sampling times. In the case of photon bunching, however, $g^{(2)}(0)>1$,
such that $\mathcal{F}_{c}(T_{S})$ has a positive slope for small
$T_{S}$.}

\textcolor{black}{Note that photon anti-bunching does not always correspond
to sub-Poissonian photon statistics, which has already been shown
for fluorescent photons \cite{Singh1983254}. We will show that also
for the optomechanical system, anti-bunching and sub-Poissonian statistics
do not necessarily coincide, cf. Fig. \ref{fig:fano_factor_blue_detuning_weak_driving}.
Furthermore, photon bunching need not correspond to super-Poissonian
statistics \cite{PhysRevA.41.475}.}
\begin{figure}
\includegraphics{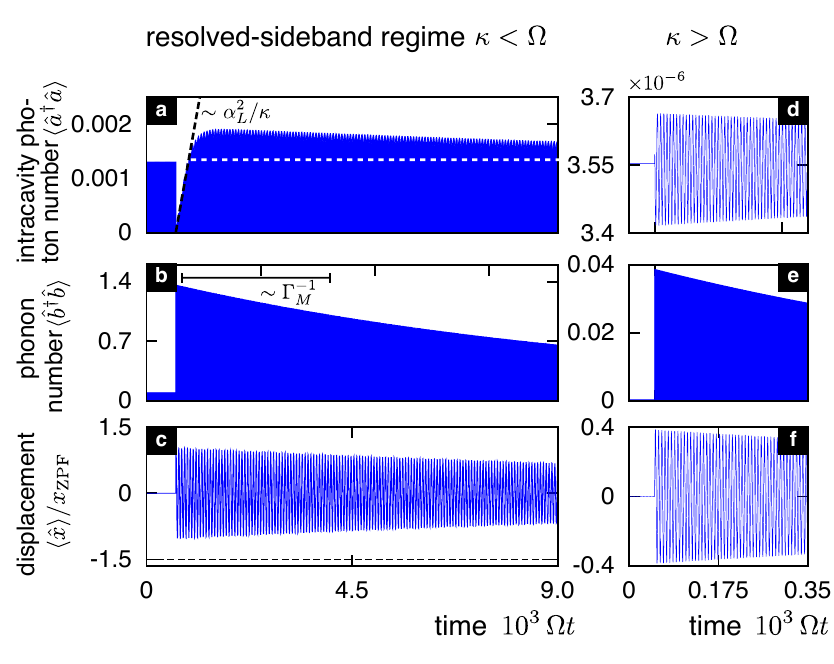}
\caption{Typical quantum jump trajectories for a weakly driven optomechanical
system. These show the expectation values for the photon number (a,d),
phonon number (b,e), and displacement (c,f), evolving conditioned
on the detection of a photon. The concomitant sudden drop in the radiation
pressure force leads to subsequent mechanical oscillations, relaxing
during the damping time $\Gamma_{{\rm M}}^{-1}$. The dashed line
in (c) corresponds to the displacement which would make the optomechanical
cavity resonant with the incoming laser. (a-c) Sideband-resolved regime
($\kappa=\Omega/8$), with $g_{0}/\kappa=4$. (d-f) Bad cavity limit
($\kappa=5\Omega$, and $g_{0}/\kappa=1/10$). {[}Parameters: Detuning
$\Delta=\Omega-g_{0}^{2}/\Omega$, coupling $g_{0}=\Omega/2$, laser
drive $\alpha_{L}=5\cdot10^{-3}\,\Omega$, mechanical damping $\Gamma_{M}=10^{-3}\,\Omega$,
photon decay $\kappa_{I}=\kappa_{O}=\kappa/2$, bath temperature $T=0${]}}
\label{fig:blue_detuning_qjt} 
\end{figure}
\subsection{Dependence of the Fano Factor on the Detector Efficiency $\eta$\label{sub:Dependence-of-F-on-eta}}
The Fano factor $\mathcal{F}_{c}(T_{S})$ depends on the detector
efficiency $\eta$. The relation
\begin{equation}
\mathcal{F}_{c}(T_{S})=\frac{\mathcal{F}_{c}^{\text{all}}(T_{S})+(\kappa-\eta\kappa_{O})/\eta\kappa_{O}}{1+(\kappa-\eta\kappa_{O})/\eta\kappa_{O}}\label{eq:fano_factor_ideal_fano_factor}
\end{equation}
connects the measured Fano factor $\mathcal{F}_{c}$ to $\mathcal{F}_{c}^{\text{all}}$.
Here, $\mathcal{F}_{c}^{\text{all}}$ is the Fano factor that would
be obtained when monitoring all decay channels with an ideal photon
detector. Thus, we can reconstruct the ideal measurement outcome even
if the output ports are observed with a non-ideal detector. Therefore,
without loss of generality, we will assume $\eta=1$ in the following.
\section{Results\label{sec:Results}}
\subsection{Weak Laser Drive\label{sub:Weak-Laser-Drive}}
\begin{figure}[t]
\includegraphics{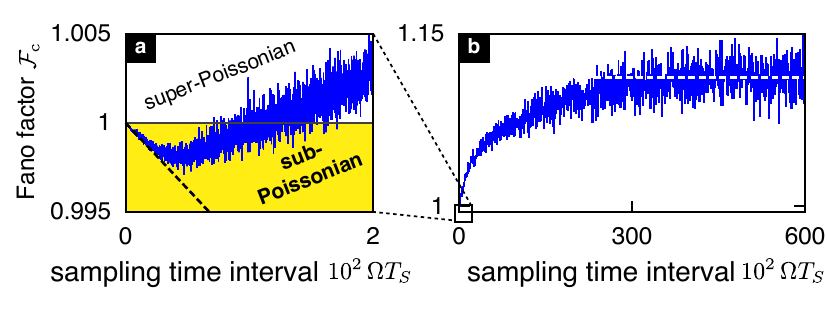}
\caption{Fano factor $\mathcal{F}_{\text{c}}(T_{S})$ for photon detection
events, as a function of the sampling time interval $T_{S}$. (a)\textbf{
}Fano factor for small sampling times, where anti-bunching and sub-Poissonian
statistics is observed ($g^{(2)}(0)<1$ and $\mathcal{F}_{\text{c}}(T_{S})<1$;
cf. Eq.~(\ref{eq:fano_factor_small_sampling_times})). (b) Fano factor
for larger sampling times $T_{S}$. We see that $\mathcal{F}_{\text{c}}(T_{S}\rightarrow\infty)>1$
saturates and indicates super-Poissonian statistics. This is a characteristical
effect of the mechanical motion induced by a photon jump (see main
text). {[}Parameters as in Fig. \ref{fig:blue_detuning_qjt}a{]}.}
\label{fig:fano_factor_blue_detuning_weak_driving} 
\end{figure}

We start by considering a weak drive such that the photon number inside
the cavity remains small, $\left\langle \hat{a}^{\dagger}\hat{a}\right\rangle \ll1$.
In contrast to previous results \cite{RablPhoton2011}, we are able
to discuss the photon correlations $g^{(2)}(\tau)$ at arbitrary time-delays,
which is crucial to capture the long time scales induced by the mechanical
motion \cite{RotterMonitoringSingleIonsMotion2008}.

Fig.~\ref{fig:blue_detuning_qjt}a-c show a typical trajectory in
the photon-blockade regime \cite{RablPhoton2011}, where the mechanical
bath temperature is assumed to be zero. The intracavity photon number
$\langle\hat{a}^{\dagger}\hat{a}\rangle$ decreases due to a photon
jump. This behavior indicates anti-bunching, i.e. $g^{(2)}(0)<1$
\cite{UedaQuantum1990}. Thus, it is less probable to detect a second
photon right after the detection of the first photon. \textcolor{black}{It
turns out that we can connect the two-photon correlation function
$g^{(2)}(\tau)$ to the intracavity photon number $\left\langle \hat{a}^{\dagger}\hat{a}\right\rangle ^{({\rm cond)}}\left(\tau\right)$
conditioned on a photon jump at $\tau=0$. This can be done via
\begin{equation}
g^{(2)}\left(\tau\right)\approx\frac{\left\langle \hat{a}^{\dagger}\hat{a}\right\rangle ^{({\rm cond)}}\left(\tau\right)}{\left\langle \hat{a}^{\dagger}\hat{a}\right\rangle \left(\tau=0^{-}\right)}\,.\label{eq:connection_g2_tau_photon_number_qj}
\end{equation}
Here $\left\langle \hat{a}^{\dagger}\hat{a}\right\rangle \left(\tau=0^{-}\right)$
denotes the photon number in any given quantum jump trajectory right
before the photon jump (white dashed line in Fig. \ref{fig:blue_detuning_qjt}a)
and $\tau$ denotes the time after the photon jump. The relation of
Eq. (\ref{eq:connection_g2_tau_photon_number_qj}) can be understood
intuitively: $\left\langle \hat{a}^{\dagger}\hat{a}\right\rangle ^{({\rm cond)}}\left(\tau\right)$
is proportional to the probability of detecting a photon conditioned
on a photon jump at $\tau=0$. As discussed in Sec. \ref{sub:PConnection_of_g2_and_Fano_factor},
this coincides with the interpretation of $g^{(2)}(\tau)$. The validity
of Eq. (\ref{eq:connection_g2_tau_photon_number_qj}) can be seen
more formally by comparing it to the quantum regression formula \cite{CarmichaelOpen1993}.
For that to hold, we have to require that photon events are rare,
which is fulfilled in the regime of weak driving: The photon number
entering the definition of the two-photon correlation function (\ref{eq:def_g2_tau})
is given by a long-time-average of the photon number gained from a
quantum jump trajectory. Thus, if the jump events are rare, the limiting
value $\left\langle \hat{a}^{\dagger}\hat{a}\right\rangle \left(\tau=0^{-}\right)$
is assumed most of the time and it will be a good approximation for
the long-time average $\left\langle \hat{a}^{\dagger}\hat{a}\right\rangle $
of the photon number. Additionally, the two-photon correlation function
$g^{(2)}(\tau)$ can be reconstructed as an ensemble average over
many quantum jump trajectories conditioned on a photon jump at $\tau=0$.
Thus, the trajectory depicted in Fig. \ref{fig:blue_detuning_qjt}
will be a valid approximation of the ensemble average when $\tau$
is smaller than the average waiting time between photon jumps. }

Let us now focus on how the photon detection influences the mechanical
degree of freedom: Before observing the single photon, the average
radiation pressure force has been very small, since $F_{\text{ra}d}\sim\left\langle \hat{a}^{\dagger}\hat{a}\right\rangle $
and $\left\langle \hat{a}^{\dagger}\hat{a}\right\rangle \ll1$. Now
that we know that a photon is detected, however, we deduce that this
single photon must have exerted a significant radiation pressure force
prior to leaving the cavity. The quantum jump formalism takes this
into account in the following way: Upon photon detection, the expectation
for the mechanical displacement is updated to reflect the value it
must have had due to this relatively large single-photon force, which
is much larger than the \emph{average} force (sudden jump of $\left\langle \hat{x}\right\rangle $
in Fig.~\ref{fig:blue_detuning_qjt}c towards a greater displacement).
This reflects photon-phonon entanglement in the state prior to photon
detection. After the photon has left the cavity, however, the radiation
pressure force becomes very small, since the photon number expectation
value $\left\langle \hat{a}^{\dagger}\hat{a}\right\rangle $ now has
dropped close to zero. Thus, the displaced mechanical resonator starts
to oscillate. This subsequent oscillation decays on the mechanical
damping time scale. The photon number itself (and, hence, $g^{(2)}\left(\tau\right)$)
quickly increases again due to the laser driving and then, much more
slowly, settles to its limiting value, on a time scale set by $\Gamma_{{\rm M}}^{-1}$.

In the following, we focus on the effects of this backaction onto
the two-photon correlations for time delays $\tau\not=0$. As shown
in Fig.~\ref{fig:blue_detuning_qjt}a, where blue detuning $\Delta>0$
is chosen, the photon number temporarily exceeds its limiting value
before settling down. In terms of the two-photon correlation function
and Eq. (\ref{eq:connection_g2_tau_photon_number_qj}), this means
that $g^{(2)}(\tau)>1$. This is because the mechanical oscillations
bring the cavity closer to resonance. Thus, even though the photon
statistics at $\tau=0$ is anti-bunched, there is the chance of observing
super-Poissonian statistics for longer observation times. This can
in fact happen, as we will show now. It should be noted that this
is another effect where the mechanical motion is crucially important.

As already discussed above, the photon statistics can also be quantified
by the Fano factor $\mathcal{F}_{{\rm c}}(T_{S})$ , which is shown
in Fig.~\ref{fig:fano_factor_blue_detuning_weak_driving}. For the
bad cavity case (Fig.~\ref{fig:blue_detuning_qjt}d-f), the Fano
factor would be close to the value expected for a simple coherent
laser beam, $\mathcal{F}_{\text{c}}(\infty)\approx1$. This is due
to the fact that the photon number conditioned on a photon jump remains
very close to its limiting value, with small oscillations (cf. Fig.
\ref{fig:blue_detuning_qjt}d). Thus, $g^{(2)}(\tau)$ will oscillate
around the coherent value $1$ as well, such that $\mathcal{F}_{\text{c}}(\infty)\approx1$,
cf. Eq. (\ref{eq:fano_factor_small_sampling_times}). In contrast,
for the strongly coupled and sideband-resolved optomechanical system
of Fig.~\ref{fig:blue_detuning_qjt}a-c, the Fano factor shows pronounced
sub-Poissonian statistics at short times ($\mathcal{F}_{\text{c}}<1$),
since $g^{(2)}(\tau)<1$ in this case, and super-Poissonian ($\mathcal{F}_{\text{c}}>1$)
statistics for long times, up to $T_{S}\rightarrow\infty$. Thus,
the full counting statistics indicates that typically groups of more
than one photon are detected during a single transmission process
(before the system settles down again), even if the photons are anti-bunched
at short times.

\textcolor{black}{Let us now briefly discuss the influence of the
mechanical bath temperature $T$ onto the resulting photon statistics.
To that end, we consider finite temperatures $T$. Note that in our
numerical simulations, we choose average thermal phonon numbers $n_{\text{th}}(T)\lesssim2$
to keep the Hilbert space manageable. These temperatures can be reached
by cryogenic cooling if the mechanical frequency is large enough.
We find that the photons remain anti-bunched even for these temperatures,
i.e. $g^{(2)}(0)<1$. For the parameters discussed here, the Fano
factor in the long-time limit $\mathcal{F}_{c}(\infty)$ decreases
with increasing bath temperature. However, the photon statistics remains
super-Poissonian, i.e. $\mathcal{F}_{c}(\infty)>1$.}

In conclusion, we emphasize that the sign of the photon statistics
(super-Poissonian vs. sub-Poissonian statistics) can depend on the
observation time interval. Thus, $\mathcal{F}_{{\rm c}}(T_{S})$ has
to be analyzed for all times $T_{S}$ to capture the full impact of
the optomechanical dynamics on the photon statistics. Additionally,
it turns out that these findings are robust against thermal fluctuations
(with $n_{\text{th}}\lesssim2$). 
\begin{figure}[t]
\includegraphics{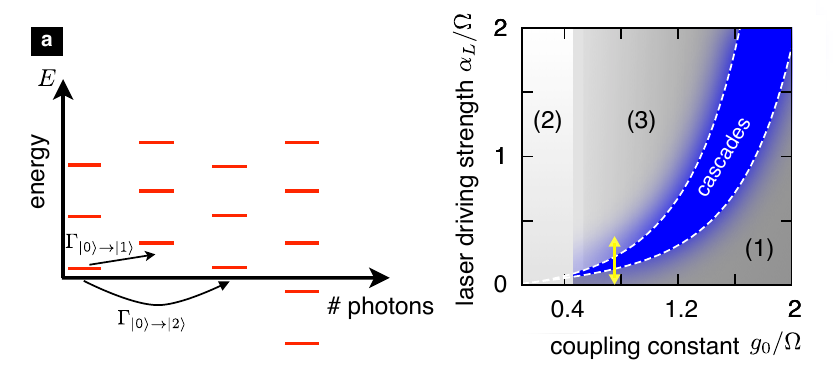}
\caption{(a) Optomechanical level scheme for a choice of red detuning that
enables the two-photon transition. (b) Map of parameters $\alpha_{L}/\Omega$
and $g_{0}/\Omega$, for which a cascade behavior is expected (dark
blue region). Note that in regions (1), (2) and (3) the first, second
and third inequality of Eq.~(\ref{eq:virtual_trans}) is violated.
The arrow indicates the trace taken in Fig. \ref{fig:fano_factor_avalanches}.
{[}Parameters: $\Delta=-2g_{0}^{2}/\Omega$ and $g_{0}/\kappa=4${]}}
\label{fig:Level_Scheme_Avalanches_and_map_of_parameters} 
\end{figure}
\subsection{Cascade of Photon Transmission at Strong Driving\label{sub:Cascade-of-Photon}}
For strong laser driving, we observe an interesting feature: Cascades
of transmitted photons. These cascades originate from multi-photon
transitions, which appear for red detuning $\Delta=-n\cdot g_{0}^{2}/\Omega$
as shown in the level scheme of Fig. \ref{fig:Level_Scheme_Avalanches_and_map_of_parameters}a.
In this cascade regime, $n>1$ photons enter the cavity simultaneously
(involving $n-1$ virtual transitions), since the incidence of $m\not=n$
photons is precluded by energy conservation due to the optomechanically
induced photon nonlinearity. To favor the virtual (but resonant) $|0\rangle\rightarrow|n\rangle$
transition over the non-virtual (but off-resonant) $|0\rangle\rightarrow|1\rangle$
transition, we impose the inequalities
\begin{equation}
\Gamma_{|0\rangle\rightarrow|n\rangle}>\Gamma_{|0\rangle\rightarrow|1\rangle}\,,\,\kappa\ll|\Delta|\,,\,\text{ and }\,\kappa\gg\Gamma_{|0\rangle\rightarrow|n\rangle}\,.\label{eq:virtual_trans}
\end{equation}
The second and third inequality of Eq. (\ref{eq:virtual_trans}) ensure
that the intermediate transitions of the resonant transition are virtual,
and to prevent heating of the resonator. The transition rates read
\begin{equation}
\Gamma_{|0\rangle\rightarrow|1\rangle}=\alpha_{L}^{2}\frac{\kappa\cdot e^{-\left(g_{0}/\Omega\right)^{2}}}{(n-1)^{2}g_{0}^{4}/\Omega^{2}+\kappa^{2}/4}
\end{equation}
and
\begin{equation}
\Gamma_{|0\rangle\rightarrow|n\rangle}=\frac{4\alpha_{L}^{2n}}{\kappa}\left(\frac{\Omega}{g_{0}^{2}}\right)^{2(n-1)}\frac{e^{-n\left(g_{0}/\Omega\right)^{2}}}{[(n-1)!]^{3}}\,.
\end{equation}
Upon increasing the laser drive, at some point the $n$-photon transition
will start to be favored over other processes. Neglecting the mechanical
motion for a moment, we expect that these $n$ photons decay out of
the cavity on a timescale $\sim\kappa^{-1}$, leading to a photon
cascade. In this case, we would naively expect $\mathcal{F}_{\text{c}}^{\text{all}}(\infty)\approx n$.
Since we observe the output port only, we expect $\mathcal{F}_{c}(\infty)\approx(n+1)/2$
for a symmetrical cavity ($\kappa_{I}=\kappa_{O}$), cf. Eq. (\ref{eq:fano_factor_ideal_fano_factor}).
Thus, the Fano factor has to increase when the multi-photon transition
starts to become important, regardless of the detector efficiency.
\begin{figure}
\includegraphics{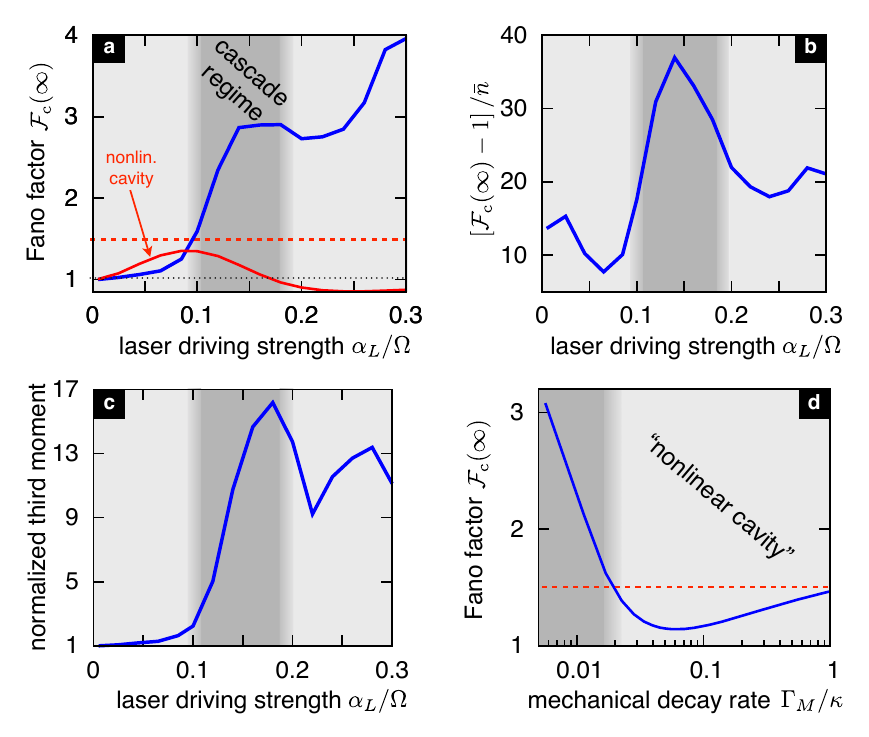}
\caption{Photon statistics in the {}``photon cascade regime''. (a) Fano factor
$\mathcal{F}_{\text{c}}(\infty)$ as a function of laser drive. The
red, dashed line represents the naive expectation $\mathcal{F}_{c}(\infty)=(n+1)/2(=1.5)$.
In the highlighted region, the multi-photon transition is favored
(cf.~Fig.~\ref{fig:Level_Scheme_Avalanches_and_map_of_parameters}b).
At the onset of this regime, the Fano factor $\mathcal{F}_{\text{c}}$
rises beyond the naive expectation. Red curve: $\mathcal{F}_{c}(\infty)$
for a non-linear cavity (Kerr nonlinearity) and same parameters. (b)
Rescaled deviation from Poisson statistics. (c) Normalized third moment,
$\left\langle (N-\left\langle N\right\rangle )^{3}\right\rangle /\left\langle N\right\rangle $.
(d) Fano factor as a function of mechanical decay rate $\Gamma_{M}$,
approaching the value of a nonlinear cavity for large $\Gamma_{M}$.
{[}Parameters (a)-(c): $\Delta=-2g_{0}^{2}/\Omega$, $g_{0}=\Omega/\sqrt{2}$,
$\kappa=g_{0}/4=\Omega/(4\sqrt{2}$), $\Gamma_{M}=10^{-3}\,\Omega$,
$T=0$. (d) Same parameters, but $\alpha_{L}=0.15\,\Omega${]}.}
\label{fig:fano_factor_avalanches} 
\end{figure}

In the following, we will focus on single-photon strong coupling $g_{0}\gtrsim\kappa$.
When a specific $n$-photon resonance is chosen and for a fixed ratio
$g_{0}/\kappa$, the laser driving strength $\alpha_{L}/\Omega$ and
the coupling strength $g_{0}/\Omega$ is varied. A map of these two
parameters is shown in Fig.~\ref{fig:Level_Scheme_Avalanches_and_map_of_parameters}b,
where the cascade regime is highlighted.

\begin{figure}
\includegraphics{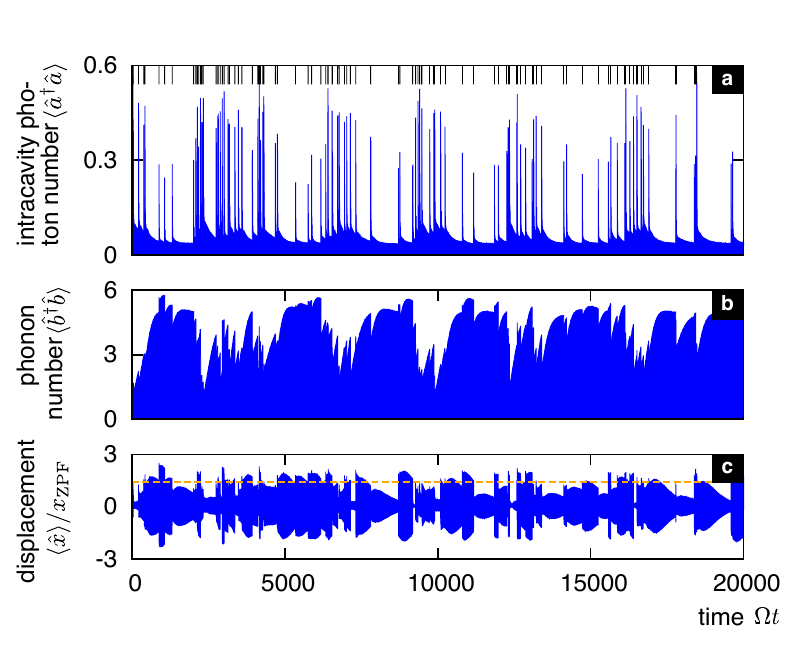}
\caption{Typical quantum jump trajectory in the cascade regime. The pattern
of vertical lines in (a) indicates the actual photon jump events,
where bunching is clearly observable. (b) Phonon number expectation
value. If a photon jump occurs, the mechanical state gets updated,
leading to an instantaneous change of the phonon number. This reflects
photon-phonon entanglement. The orange, dashed line in (c) corresponds
to the displacement that would make the cavity resonant with the incoming
laser. {[}Parameters: as in Fig.~\ref{fig:fano_factor_avalanches}a,
and $\alpha_{L}=0.15\,\Omega${]}.}
\label{fig:avalanche_trajectory} 
\end{figure}

In Fig.~\ref{fig:fano_factor_avalanches}a, the Fano factor $\mathcal{F}_{\text{c}}(\infty)$
is shown as a function of the laser drive. We see that $\mathcal{F}_{\text{c}}\left(\infty\right)\approx1$
for weak laser driving (Poisson statistics), while $\mathcal{F}_{\text{c}}(\infty)$
increases rapidly beyond the threshold of the laser drive predicted
by the above analysis. There, the $n$-photon transition is favored
and photons are emitted in bunches. In contrast to our naive expectation,
$\mathcal{F}_{\text{c}}(\infty)$ is larger than $(n+1)/2$ and, hence,
$\mathcal{F}_{c}^{\text{all}}(\infty)>n$. This fact is due to the
mechanical motion induced by photon detection events, cf. Fig.~\ref{fig:avalanche_trajectory},
where a typical jump trajectory in the cascade regime is displayed. 

Upon detection of a photon leaving the cavity, the expectation value
of the intracavity photon number increases, cf. Fig. \ref{fig:avalanche_trajectory}a.
This behavior is known for situations with photon bunching \cite{UedaQuantum1990,ParigiProbing2007},
and it arises because the detection of a photon is most likely due
to a previous multi-photon transition \cite{UedaQuantum1990,ParigiProbing2007}.
Thus, upon observing one photon leaving the cavity, we can now reasonably
expect that there are still more photons inside the cavity. These
remaining photons are likely to be emitted rapidly afterwards (the
rate for $n$ photons to decay is $n\kappa$). However, as discussed
above in our analysis of the weak-driving case, the mechanical resonator
gets displaced by a photon jump and starts to oscillate subsequently
(cf. Fig. \ref{fig:avalanche_trajectory}b and c). This mechanical
motion can bring the cavity into resonance (cf. Fig. \ref{fig:avalanche_trajectory}c),
which then allows for the transmission of additional photons. In this
way, cascades of more than $n$ transmitted photons become possible
(i.e. $\mathcal{F}_{c}^{\text{all}}(\infty)>n$). Note that once a
single cascade of photon transmission terminates, no photon jumps
are observed any more and the mechanical displacement is able to relax
towards its equilibrium value.

Let us now focus on even larger laser driving strength. In this case,
the third inequality of Eq.~(\ref{eq:virtual_trans}) is violated,
and the average photon and phonon numbers as well as $\mathcal{F}_{\text{c}}(\infty)$
increase even further. Additionally, no clear cascades are observable
anymore. The strong photon noise in that case can be explained as
being due to the transmission modulated stochastically by a fluctuating
mirror, whose motion has been heated. This is seen in a decrease of
the rescaled Fano factor $(\mathcal{F}_{{\rm c}}(\infty)-1)/\bar{n}$
(Fig.~\ref{fig:fano_factor_avalanches}b): \textcolor{black}{The
rescaling is introduced to be able to distinguish two different regimes
in which the Fano factor tends to be large but where the fluctuations
are of qualitatively different origin. In the photon cascade regime,
the increase of the Fano factor is due to large and long-time intracavity
photon correlations. Since the rescaled Fano factor is maximal in
this regime, the time-averaged intracavity photon correlations are
strongest, cf. Eq. (\ref{eq:fano_factor_small_sampling_times}). As
we have seen, the Fano factor increases further for even larger laser
driving strengths. By inspecting the rescaled Fano factor we observe,
however, that the time-averaged intracavity photon correlations decrease.
Thus, the Fano factor in this case increases further due to an increased
photon transmission rate and not because of large intracavity photon
correlations. }

Note that the statistical data obtained from the quantum jump trajectories
also provide access to higher moments. The third moment (Fig.~\ref{fig:fano_factor_avalanches}c),
for example, displays a maximum in the cascade regime studied here.

Let us now discuss two other pieces of evidence which show that the
enhanced cascade of photon transmission is due to the presence of
the mechanical resonator. First, we compare our optomechanical system
to a non-linear cavity (Kerr nonlinearity). There, we find that $n$-photon
transitions are observable as well, cf. red curve in Fig. \ref{fig:fano_factor_avalanches}a.
However, in contrast to the optomechanical system, the Fano factor
in that case does not exceed the naively expected value $\mathcal{F}_{c}(\infty)\approx(n+1)/2$. 

Second, as another indication of the strong mechanical signature in
the photon correlations, we note that the Fano factor depends strongly
on the mechanical damping rate (Fig. \ref{fig:fano_factor_avalanches}d).
For smaller damping, the mechanical resonator oscillates more often
through the cavity resonance such that even more photons are transmitted,
and the Fano factor increases. On the other hand, by choosing $\Gamma_{M}\sim\kappa$,
the limit of a non-linear cavity can be achieved. Experimentally,
$\Gamma_{M}$ could be tuned by making use of the optomechanical cooling
effect \cite{MarquardtCooling2007,Wilson-RaeCooling2007} using a
second optical mode. 

\begin{figure}[t]
\includegraphics{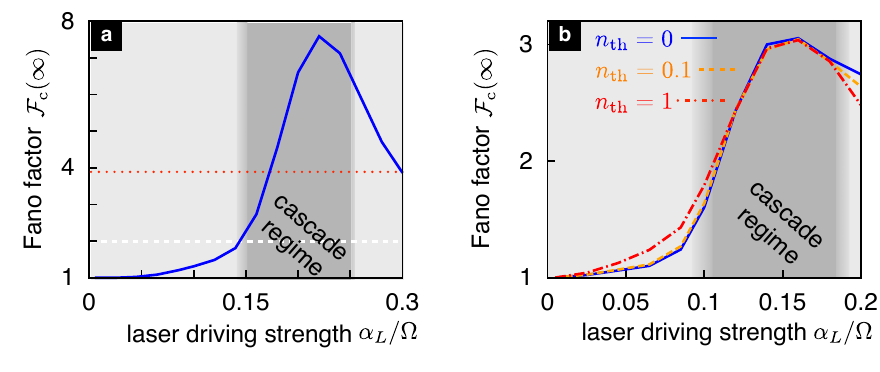}
\caption{\textcolor{black}{(a) Fano factor $\mathcal{F}_{c}(\infty)$ for a
higher resonance $(n=3)$ showing photon cascades. The white dashed
line shows the naive expectation $\mathcal{F}_{c}(\infty)=(n+1)/2$.
The red, dotted line indicates the maximum value of $\mathcal{F}_{c}(\infty)$
which is reached by a non-linear cavity (Kerr nonlinearity) with the
same parameters. (b) Fano factor $\mathcal{F}_{c}(\infty)$ for $n=2$
and different temperatures. {[}Parameters: (a) $n=3$, $\Delta=-3g_{0}^{2}/\Omega_{M}$,
$g_{0}=\Omega/2$, $g_{0}/\kappa=4$, $T=0$ (b) Same parameters as
in Fig. \ref{fig:fano_factor_avalanches}, but $n_{\text{th }}=0$
(solid blue line), $n_{\text{th}}=0.1$ (dashed orange line) and $n_{\text{th}}=1$
(dash-dotted red line){]}.}}
\label{fig:7}
\end{figure}

\textcolor{black}{Let us now briefly discuss a higher resonance condition,
with $n>2$, cf. Fig. \ref{fig:7}a. In general, the minimum laser
driving strength needed to generate photon cascades increases with
$n$. This is because $n$ photons enter the cavity simultaneously
via $n-1$ virtual transitions. Furthermore, we observe the same enhanced
cascade behavior as for $n=2$: First, we see that the Fano factor
rises strongly for sufficiently large laser driving, as it was the
case for $n=2$. Second, as already discussed before, we can compare
the optomechanical system to a non-linear cavity (Kerr nonlinearity).
We observe that the Fano factor for the Kerr cavity lies below the
optomechanical Fano factor. Thus, we infer that photon transmission
is strongly enhanced by the mechanical resonator for $n>2$ as well.}

\textcolor{black}{Finally, in the following, we will discuss the influence
of the mechanical bath temperature $T$ onto the photon cascades (taking
the case $n=2$ for concreteness). We consider the Fano factor $\mathcal{F}_{c}(\infty)$
for different laser driving strengths and different temperatures,
cf. Fig. \ref{fig:7}b. Let us first focus on laser driving strengths
$\alpha_{L}$ lying below the cascade regime. There, we find that
the Fano factor increases with increasing temperature for fixed $\alpha_{L}$.
This is because the thermal fluctuations of the position of the cantilever
allow for the stochastic transmission of photons, which leads to an
increase of $\mathcal{F}_{c}(\infty)$. In the photon cascade regime,
however, we observe that the Fano factor does not change strongly
with increasing temperature. This is because for strong laser driving
many phonons are generated by the coherent dynamics even for zero
temperature, cf. Fig. \ref{fig:avalanche_trajectory}b. Thus, in contrast
to the weak driving case discussed above, the few thermal phonons
do not affect the dynamics of the mechanical resonator considerably.}

\section{Conclusion\label{sec:Conclusion}}
We have analyzed the full statistics of photons transmitted through
an optomechanical system. The photon correlations crucially depend
on the observation time and may even change from sub-Poissonian to
super-Poissonian statistics in the long-time limit. For larger laser
driving, photon cascades may be observed. These contain clear signatures
of the strong light-mechanics coupling, and the influence of the mechanical
motion could even be tuned by varying the mechanical decay rate.

\textit{Note added. }Recently, we became aware of a related paper
by Xu, Li, and Liu \cite{XuPhotonInducedTunneling2012}.
\begin{acknowledgments}
Financial support by DARPA ORCHID, the Emmy-Noether program and the
ERC is gratefully acknowledged. A.K. thanks the HPC group at the FAU
Erlangen-Nürnberg for stimulating discussions. 
\end{acknowledgments}
\bibliographystyle{apsrev4-1}
\bibliography{references}
\end{document}